\documentclass[preprint,12pt]{elsarticle}
\biboptions{sort&compress}
\usepackage{amssymb}
\journal{Physics Letters A}

\begin{document}

\begin{frontmatter}
\title{Complex network approach for recurrence analysis of time series}

\author[pik]{Norbert Marwan}
\ead{marwan@pik-potsdam.de}
\author[pik,hu]{Jonathan F.~Donges}
\author[pik]{Yong Zou}
\author[pik,tud,opu]{Reik V.~Donner}
\author[pik,hu]{J\"urgen Kurths}

\address[pik]{Potsdam Institute for Climate Impact Research, P.O.~Box 60\,12\,03, 14412 Potsdam, Germany}
\address[hu]{Department of Physics, Humboldt University Berlin, Newtonstr.~15, 12489 Berlin, Germany}
\address[tud]{Institute for Transport and Economics, Dresden University of Technology, Andreas-Schubert-Str.~23, 01062 Dresden, Germany}
\address[opu]{Graduate School of Science, Osaka Prefecture University, 1-1 Gakuencho, Naka-ku, Sakai, 599-8531, Japan}

\begin{abstract}
We propose a novel approach for analysing time series using complex
network theory. We identify the recurrence matrix (calculated from 
time series) with the adjacency matrix of a complex network 
and apply measures for the characterisation of complex networks 
to this recurrence matrix. By using the logistic map, we 
illustrate the potential of these complex network measures 
for the detection of dynamical transitions. Finally, we apply 
the proposed approach to a marine palaeo-climate record and 
identify the subtle changes to the climate regime. 
\end{abstract}


\end{frontmatter}


\section{Introduction}
In many scientific disciplines, such as engineering, astrophysics, 
life sciences and economics, modern data analysis techniques are 
becoming increasingly popular as a means of understanding 
the underlying complex dynamics of the system. Methods for 
estimating fractal or correlation dimensions, Lyapunov 
exponents, and mutual information have been widely used
\cite{kantz97,kurths87,mandelbrot82,wolf85}. 
However most of these methods require long data series 
and in particular their uncritical application, especially 
to real-world data, may often lead to pitfalls.

In the last two decades, the method of recurrence plots has been developed 
as another approach to describe complex dynamics
\cite{marwan2007}. 
A recurrence plot (RP) is the graphical representation of a binary 
symmetric square matrix which encodes the times when two states  are in
close proximity (i.e.~neighbours in phase space).  Based on such a
recurrence matrix, a large and diverse  amount of information on the
dynamics of the system can be  extracted and statistically quantified 
(using recurrence quantification analysis, dynamical invariants, etc.). 
Meanwhile this technique has been the subject of much interest  from various
disciplines \citep{marwan2008epjst} and it has been successfully applied  to
a number of areas: the detection of dynamical transitions
\citep{trulla96,ngamga2007}  and synchronisation \citep{romano2005}, the
study of protein structures \citep{giuliani2002a,zbilut2004b} and in 
cardiac and bone health conditions \cite{marwan2002herz,marwan2007pla}, in
ecological regimes \citep{facchini2007,proulx2009},  economical dynamics
\citep{kyrtsou2005,bigdeli2009}, in chemical reactions
\citep{castellini2004} and to monitor  mechanical behaviour and damages in
engineering \citep{nichols2006,sen2008}, to name a few.  It is important to
emphasise that  recurrence plot based techniques are even useful for the
analysis of short and non-stationary data, which often presents a critical
issue when studying real world data. The last few years have witnessed great
progress in the development of RP-based approaches for the analysis of
complex systems
\citep{marwan2007,thiel2004b,marwan2005,romano2007,robinson2009}.

During the last decade, complex networks have become rather popular for
the analysis of complex and, in particular, spatially extended systems
\cite{watts1998,strogatz2001,boccaletti2006,arenas2008}. Local and global properties 
(statistical measures) of complex networks are helpful to understand
complex interrelations and information flow between different components
in extended systems, such as social, computer or neural networks \cite{watts1998}, 
food webs, transportation networks, power grids \citep{albert2004}, or even in the 
global climate system \citep{donges2009epjst}. The basis of complex network
analysis is the adjacency matrix, representing the links between the nodes
of the network. Like the recurrence matrix, the adjacency matrix is also 
square, binary, and symmetric (in the case of an unweighted and undirected network).

In fact, the recurrence matrix and the adjacency matrix exhibit a strong
analogy: a recurrence matrix represents neighbours in phase space 
and an adjacency matrix represents links in a network; both matrices 
embody a pair-wise test of all components (phase space vectors 
resp.~nodes). Therefore, we might well proceed to
explore further analogies even in the 
statistical analysis of both the recurrence and the adjacency matrix.

Quantitative descriptors of RPs have been first
introduced in a heuristic way in order to distinguish different
appearances of RPs \cite{marwan2008epjst}. We may also consider 
to apply measures of complex network theory to a RP in order
to quantify the RP's structure and the corresponding 
topology of the underlying phase space trajectory. In this 
(more heuristic) sense, it is actually not necessary to consider 
the phase space trajectory as a network.

Recently, the very first steps in the direction of bridging complex network
theory and recurrence analysis have been reported 
\cite{zhang2006,xu2008}. In these works, the local properties of
phase space trajectories have been studied using complex network measures.
Zhang et al.~suggested using cycles of the phase space trajectory
as nodes and considering a link when two cycles are rather similar 
\cite{zhang2006,zhang2008netw}. The resulting adjacency matrix can be in fact 
interpreted as a special recurrence matrix. The recurrence criterion here
is the matching of two cycles. A complementary approach was suggested
by Xu et al.~who studied the structural shape of the direct neighbourhood
of the phase space trajectory by a motif classification \cite{xu2008}. 
The adjacency matrix of the underlying network corresponds to the
recurrence matrix, using the recurrence criterion of a fixed number
of neighbours (instead of the more often used fixed size of
the neighbourhood \cite{marwan2007}). 

Other approaches for the study of time series by a complex network
analysis suggested using linear correlations \cite{yang2008nw} or another 
certain condition on the time series amplitudes (``visibility'') \cite{lacasa2008}.

In this letter, we demonstrate that the recurrence matrix (analogously
to \cite{xu2008}) can be considered
as the adjacency matrix of an undirected, unweighted network, allowing
us to study time series using a complex network approach. This
ansatz on creating complex network is more natural and simple than the
various suggested approaches \cite{zhang2008netw,yang2008nw,lacasa2008}.
Complex network statistics is helpful to characterise the local
and global properties of a network. We propose using
these complex network measures for a quantitative description
of recurrence matrices. By applying these measures,
we obtain additional information from the recurrence plots,
which can be used for characterising the dynamics of the underlying
process. 
We give an interpretation of this approach in the 
context of the dynamics of a phase space trajectory. 
Nevertheless, many of
these measures neither have an analogue in traditional RQA nor in
nonlinear time series analysis in a wider sense, and hence, open up
new perspectives for the quantitative analysis of dynamical
systems. We illustrate our approach with 
a prototypical model system and a real-world example from the Earth 
sciences.

\section{Recurrence plots and complex networks}

A recurrence plot is a representation of recurrent states of a dynamical
system in its $m$-dimensional phase space. It is a pair-wise test of all 
phase space vectors $\vec{x}_i$ ($i=1,\ldots,N, \vec{x} \in \mathcal{R}^m$) 
among each other, whether or not they are close:
\begin{equation}
R_{i,j} = \Theta\bigl(\varepsilon - d(\vec{x}_i,\vec{x}_j)\bigr),
\end{equation}
with $\Theta(\cdot)$ being the Heaviside function and $\varepsilon$
a threshold for proximity \cite{marwan2007}.
The closeness $d(\vec{x}_i,\vec{x}_j)$ can be measured in 
different ways, by using, e.g., spatial distance, string metric, or
local rank order \cite{marwan2007,bandt2008}.
Mostly, a spatial distance is considered in terms of maximum or Euclidean norm
$d(\vec{x}_i,\vec{x}_j) = \|\vec{x}_i-\vec{x}_j \|$. 
The binary recurrence matrix $\mathbf{R}$ contains the
value one for all close pairs $\|\vec{x}_i-\vec{x}_j \| < \varepsilon$.
A phase space trajectory can be reconstructed from a 
time series $\{u_i\}_{i=1}^N$ by time delay embedding \cite{packard80}
\begin{equation}
\vec{x}_i = (u_i, u_{i+\tau}, \ldots, u_{i+\tau (m-1)}),
\end{equation}
where $m$ is the embedding dimension and $\tau$ is the 
delay.

The resulting matrix $\mathbf{R}$ exhibits the
line of identity (the main diagonal) $R_{i,i} = 1$. Using 
a spatial distance as the recurrence criterion, the RP
is symmetric. Small-scale features in a RP can be observed
in terms of diagonal and vertical lines. The presence of such lines reflects
the dynamics of the system and is related to divergence
(Lyapunov exponents) or intermittency \cite{trulla96,marwan2002herz,robinson2009}.
Following a heuristic approach, a quantitative description of RPs
based on these line structures was introduced and is known as
recurrence quantification analysis (RQA) \cite{marwan2008epjst}.
We use the following two RQA measures (a comparable study
using other measures can be found in \cite{marwan2002herz}).

Similarly evolving epochs of the phase space trajectory cause
diagonal structures parallel to the main diagonal. The length
of such diagonal line structures depends on the predictability
and, hence, the dynamics of 
the system (periodic, chaotic, stochastic). Therefore, the
distribution $P(l)$ of diagonal line lengths $l$ can be used for characterising
the system's dynamics. Several RQA measures are based on $P(l)$.
However, here we focus only on the maximal diagonal line length,
\begin{equation}
L_{\max} = \max \left(\left\{l_i\right\}_{i=1}^{N_l}\right),
\end{equation}
where $N_l=\sum_{l \geq l_{\min}} P(l)$ is the total number of diagonal lines.
For the definition of a diagonal line, we use a minimal length $l_{\min}$ 
\cite{marwan2007}. The length of diagonal lines corresponds
to the predictability time. In particular, the cumulative distribution of
the line lengths can be used to estimate the correlation entropy $K_2$,
i.e.~the lower limit of the sum of the positive Lyapunov exponents
\cite{marwan2007}. Hence the inverse of $L_{\max}$
gives a first rough impression of the divergence
(Lyapunov exponent) of the system.

Slowly changing states, as occurring during laminar phases 
(intermittency), result in vertical structures in the RP. Therefore,
the distribution $P(v)$ of vertical line lengths $v$ can be used to
quantify laminar phases occurring in a system. A useful measure for
quantifying such laminar phases is the ratio of 
recurrence points forming vertical structures to all recurrence points, 
\begin{equation}
LAM = \frac{ \sum_{v=v_{\min}}^N v\, P(v) }{ \sum_{v=1}^N v\, P(v)},
\end{equation}
which is called laminarity \cite{marwan2007}. 

Now let us consider the phase space vectors as nodes of a
network and identify recurrences with links. An undirected and
unweighted network is
represented by the binary adjacency matrix $\mathbf{A}$, where 
a connection between nodes $i$ and $j$ is marked as $A_{i,j} = 1$.
Excluding self-loops, we obtain $\mathbf{A}$ from the RP by removing 
the identity matrix,
\begin{equation}
A_{i,j} = R_{i,j} - \delta_{i,j},
\end{equation}
where $\delta_{i,j}$ is the Kronecker delta. Removing the identity
is not a problem, as this is also done in the analysis of
RPs (e.g.~when considering a Theiler window for RQA) \cite{marwan2007}. 
Henceforth, we regard the recurrence matrix (with applied Theiler
window) to be an adjacency matrix. Note that this way each 
state vector in phase space is represented by
one distinct node; even if two time-separated state 
vectors are identical, they are identified with two different nodes (which are 
perfect neighbours and therefore linked independently of the 
threshold $\varepsilon$; such nodes are also called twins \cite{romano2009}).



Local and global properties of a network are statistically
described by complex network measures based on the adjacency 
matrix $A_{i,j}$. To illustrate the potential of a recurrence
analysis by means of complex network theory, 
we consider several global and local
network measures that are well studied in 
literature \cite{boccaletti2006}. 


The complex network approach allows to harness the
distributions of locally defined measures for
the quantification of recurrence matrices. 
In this work, we particularly consider the degree centrality
\begin{equation} 
k_v = \sum_{i=1}^{N} A_{v,i}, 
\end{equation}
giving the number of neighbours of node $v$. 
The degree centrality is hence locally
defined and depends only on local adjacency information in a
topological sense. $k_v$ is proportional to the local
recurrence rate, as seen from a RQA point of view. Hence, it may 
be considered as a measure for the local phase space density. We 
refer to its frequency distribution $P(k)$ as the degree distribution.


Furthermore, a complex network may be globally described by its link
density, clustering coefficient and average path length. While
the normalised averaged degree centrality, called link density,
\begin{equation}\label{eq_linkdensity}
\rho = \frac{1}{N(N-1)}\sum_{i,j=1}^N A_{i,j} 
\end{equation}
corresponds to the global recurrence rate, the latter
two measures allow quantifying novel aspects of recurrence
matrices. The clustering coefficient $\mathcal{C} = \sum_v C_v/N$
gives the probability that two neighbours (i.e.~recurrences) 
of any state are also neighbours \cite{watts1998}. 
It is obtained as the average of the local clustering coefficient
\begin{equation} 
C_v = \frac{\sum_{i,j=1}^N A_{v,i}A_{i,j}A_{j,v}}{k_v(k_v-1)}.
\end{equation}

The average length of shortest paths between all
pairs of nodes is given by the average path length
\begin{equation} 
\mathcal{L} = \frac{1}{N(N-1)} \sum_{i,j=1}^N d_{i,j}, 
\end{equation}
where the length of a shortest path $d_{i,j}$ is
defined as the minimum number of links that have to be crossed
to travel from node $i$ to node $j$ \cite{boccaletti2006}.
Disconnected pairs of nodes are not included in the average
(for a detailed discussion see \cite{newman2003}). Note that it
is particularly interesting to study clustering coefficient and
average path length in unison, since both measures taken
together allow to characterise ``small-world" behaviour in
complex networks \cite{watts1998}. In a separate study,
we link the properties of a complex network with the topology
of a phase space representation of a dynamical system in more
detail \cite{donner2009netw}.
In particular, a complex network based on a recurrence 
plot usually does not exhibit the small-world feature, 
since graph distances are directly related to distances 
in phase space (i.e.~there are no ``shortcuts'' between 
distant nodes).


\section{Application to logistic map}

We illustrate the potential of the proposed approach by
an analysis of the logistic map
\begin{equation}\label{log_eq}
x_{i+1}=a\,x_{i}\left( 1-x_{i}\right),
\end{equation}
especially within the interesting range of the control parameter $a\in
[3.5,4]$ with a step size of $\Delta a=0.0005$. In the
analysed range of $a$, various dynamic regimes and transitions between them can be found,
e.\,g., accumulation points, periodic and chaotic states, band
merging points, period doublings, inner and outer crises
\cite{collet80,oblow88,wackerbauer94}. This system has been
used to illustrate the capabilities of RQA. It was shown that diagonal
line based RQA measures are able to detect chaos-order transitions
\cite{trulla96} and vertical line based measures even detect
chaos-chaos transitions \cite{marwan2002herz}.

Since Eq.~(\ref{log_eq}) is a one-dimensional map,
we compute the RP without embedding. For the study of transitions,
it is recommended to use a recurrence threshold $\varepsilon$ 
preserving a fixed recurrence rate, say 5\%. However, in the special
case of the logistic map, such approach leads to problems within
the periodic windows. In these windows the states are rapidly
alternating between subsequent time steps, leading to a high
recurrence rate (larger than 25\%). Therefore, a threshold for
preserving 5\% recurrence rate does not exist and, hence,
we cannot compute the network measures within the periodic
windows. To circumvent this, we will use a fixed recurrence 
threshold $\varepsilon$ for the example of the logistic map
(for the real world example in Sect.~\ref{sec:appl}, we will use
the preferred criteria of constant recurrence rate). 
The threshold $\varepsilon$ is selected to be 5\% of the standard 
deviation $\sigma$ of the time series \cite{schinkel2008}.

For periodic dynamics, band merging, laminar states
(cross points of supertrack functions, cf.~\cite{marwan2002herz}),
and outer crisis, we investigate the network measures
in more detail (Tab.~\ref{tab_param}). The band merging corresponds to 
intermittency, the inner crisis to certain chaos-chaos transition and the outer 
crisis to fully chaotic dynamics (all these transitions are
chaos-chaos transitions).

For these four cases, we compute a
time series of length $N=10,000$. In order to exclude transient responses
we remove the leading $1,000$ values from the data series in the following
analysis (thus we use $9,000$ values).

\begin{table}[htb]
\caption{Control parameter, RQA and network measures for different dynamical
regimes of the logistic map (RP parameter: $m=1$, $\varepsilon=0.05\sigma$).}\label{tab_param}
\centering \begin{tabular}{lrrrrrr}
\hline
Regime				&$a$	&$L_{\max}$	&$LAM$	&$\mathcal{L}$	&$\mathcal{C}$	&$\rho$	\\
\hline
\hline
period-3 			&3.830	&8996	&0		&1		&1		&0.333\\
band merging		&3.679	&49		&0.42	&22.8	&0.83	&0.050\\
laminar 			&3.791	&39		&0.12	&23.3	&0.79	&0.040\\
outer crisis		&4.000	&23		&0.20	&23.6	&0.82	&0.046\\
\hline
\end{tabular}
\end{table}


%
%
%
\begin{figure}[htbp]
\centering \includegraphics[width=\textwidth]{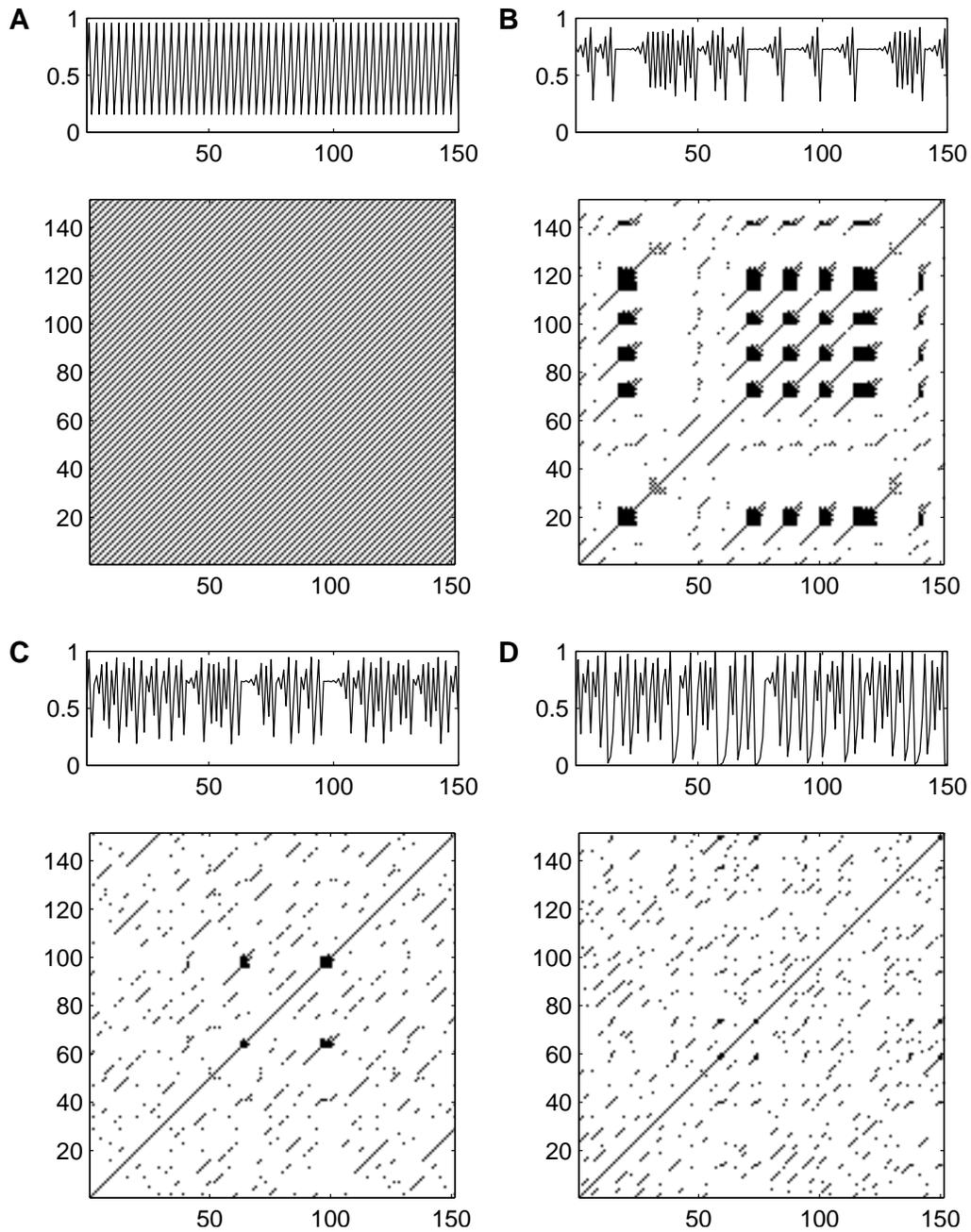}
\caption{Recurrence plots for different dynamical regimes
of the logistic map: (A) period-3 dynamics, $a=3.830$;
(B) band merging, $a=3.679$; (C) laminar states, $a=3.791$;
and (D) outer crisis, $a=4$ (RP parameters: 
$m=1$, $\varepsilon=0.05\sigma$).}\label{fig_rps}
\end{figure} 

The recurrence plots for the 
four different dynamical regimes exhibit different
typical characteristics of regular, laminar and chaotic
dynamics (Fig.~\ref{fig_rps}). In the periodic regime, $a=3.830$, the 
RP consists only of non-interrupted diagonal lines (Fig.~\ref{fig_rps}A). 
Their distance is 3, corresponding to the period length
of 3 for this periodic regime. At the band merging point, $a=3.679$,
the RP reveals extended clusters of recurrence points,
corresponding to many laminar phases (Fig.~\ref{fig_rps}B). 
Moreover, several diagonal lines appear, showing short epochs 
of similar evolution of the states. The RP for laminar states, 
$a=3.791$, consists also of (even though less) extended clusters, 
but possesses more 
diagonal lines (Fig.~\ref{fig_rps}C). For the outer crisis, 
$a=4$, diagonal lines appear but are shorter than those
appearing for smaller $a$ (Fig.~\ref{fig_rps}D), which is  
consistent with the Lyapunov exponent being largest for $a=4$ (with 
respect to smaller $a$).


\begin{figure}[tbp]
\centering \includegraphics[width=.72\textwidth]{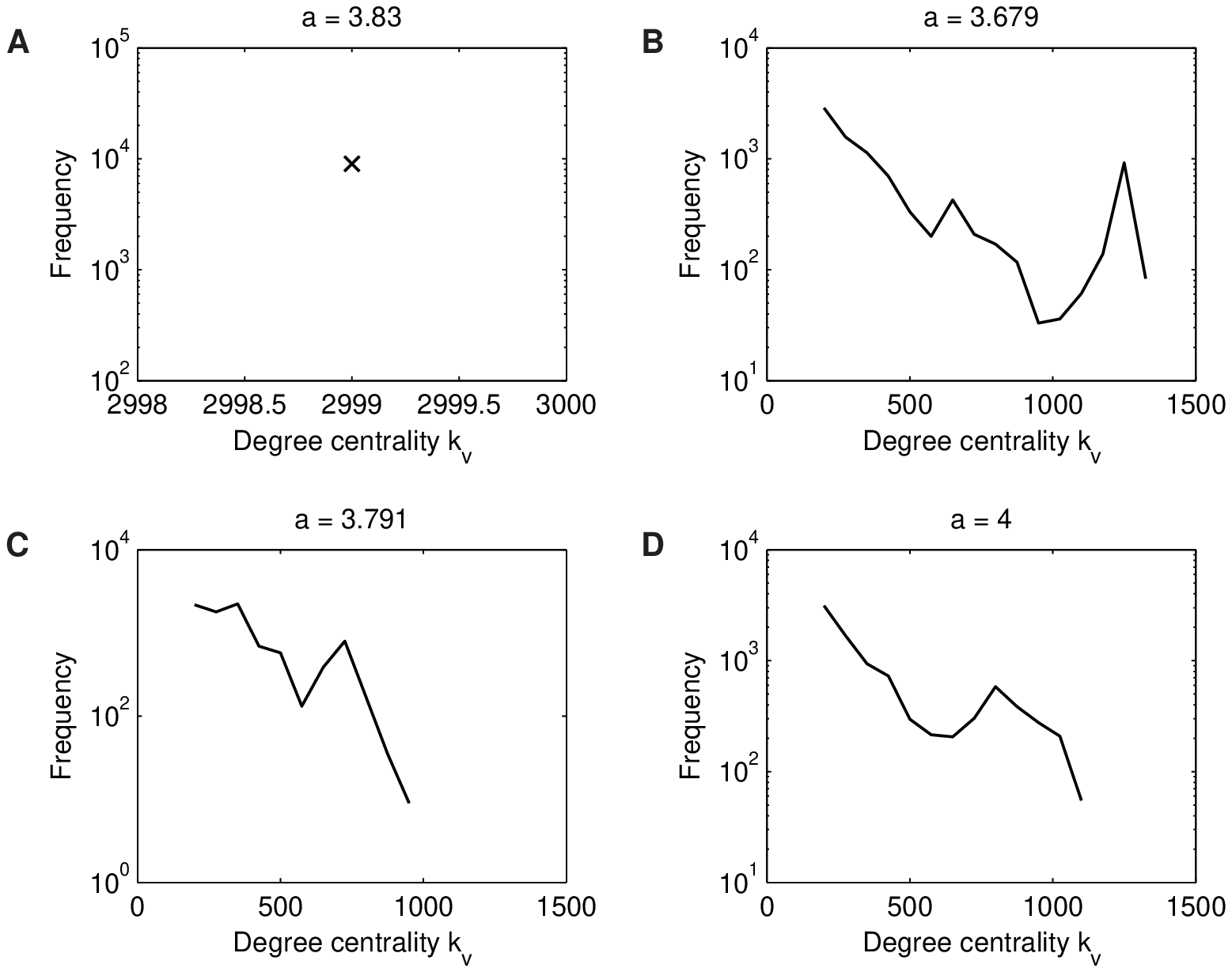}
\caption{Degree centrality distributions $P(k)$ for different 
dynamical regimes of the logistic map: (A) period-3 dynamics, $a=3.830$;
(B) band merging, $a=3.679$; (C) laminar states, $a=3.791$;
and (D) outer crisis, $a=4$ (RP parameters: 
$m=1$, $\varepsilon=0.05\sigma$).}\label{fig_rho_dist}
\end{figure} 

The two RQA measures $L_{\max}$ and
$LAM$ confirm these visual observations (Tab.~\ref{tab_param}). 
For the period-3
regime, we find the longest diagonal lines ($L_{\max} = 8996$, 
after consideration of the Theiler window \cite{theiler86}).
The maximal length of diagonal lines decreases for increasing
control parameter $a$. As expected,  
laminarity takes the highest value at the band merging point ($a=3.679$)
with $LAM = 0.42$, but is lowest for the period-3 regime, 
$LAM = 0$. At intersections of supertrack functions, the
laminarity is slightly increased ($LAM = 0.12$), and at the outer 
crisis the intermittency increases apparently ($LAM = 0.20$).

The complex network measures also highlight differences
in the topological structure of these dynamical
regimes (Tab.~\ref{tab_param}).

In the period-3 regime ($a=3.830$), the observed values jump between three distinct states.
These three states are isolated in phase space and are not considered
to be neighbours (in the sense of the recurrence definition). Therefore,
in the sense of a complex network, we have three disconnected components
where each component contains a fully connected network (because all
the nodes in each component represent the same state in phase space). 
The average shortest
path length between nodes (i.e.~states) should therefore be one, 
and the clustering is perfect.
The average path length $\mathcal{L}$ derived from the corresponding
RP has indeed the smallest possible value ($\mathcal{L} = 1$),
and the clustering coefficient $\mathcal{C}$ takes its largest
possible value ($\mathcal{C} = 1$). The degree centrality $k_v$
takes only one value: 2999 (Fig.~\ref{fig_rho_dist}A).
This value corresponds approximately to
a third of the size of the network,
due to its partition by the period-3 cycles, which is
confirmed by the link density ($\rho = 0.333$). 

\begin{figure}[htbp]
\centering \includegraphics[width=\textwidth]{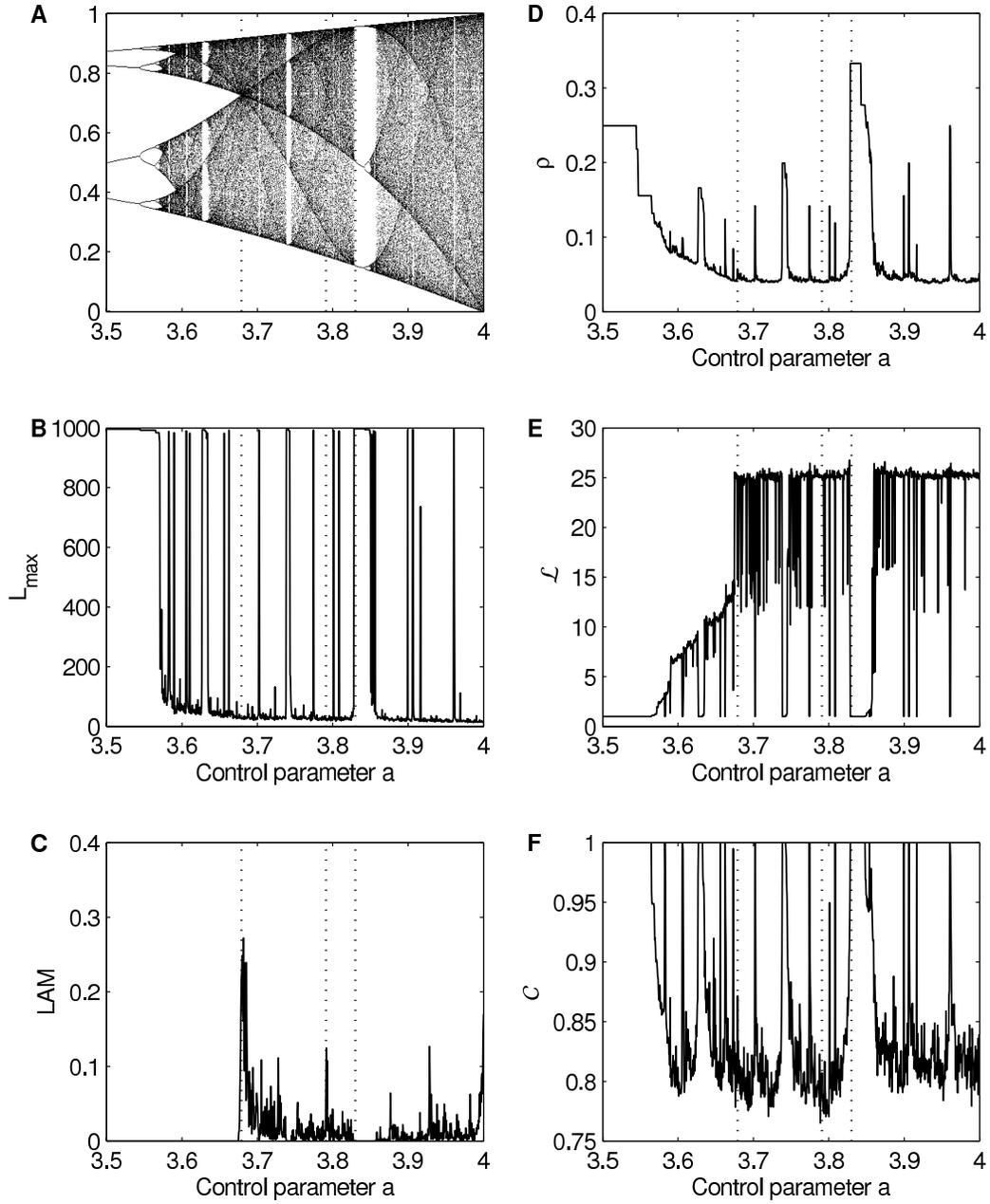}
\caption{(A) Bifurcation diagram of the logistic map.
Selected RQA measures 
(B) maximal diagonal line length $L_{\max}$ and 
(C) laminarity $LAM$, as well as complex network measures 
(D) link density $\rho$, 
(E) average path length $\mathcal{L}$, and
(F) clustering coefficient $\mathcal{C}$.
The dotted lines mark the discussed regimes at 
period-3 window ($a=3.830$), band merging
($a=3.679$), cross points of supertrack functions ($a=3.791$),
and outer crisis ($a=4$). Parameters as in Fig.~\ref{fig_rho_dist}.
}\label{fig_measures_a}
\end{figure}

For the band merging ($a=3.679$), we find $\mathcal{L} = 22.8$
and $\mathcal{C} = 0.83$. The degree distribution $P(k)$ has 
a multimodal shape (Fig.~\ref{fig_rho_dist}B), which implies that 
there are several states acting like super-nodes
(i.e.~which exhibit many links). These states lie at the
merging point of the two bands (around $x = 0.73$) and at
the upper and lower border of the state space, i.e.~in 
regions with high phase-space density (Fig.~\ref{fig_measures_a}A). 
The link density is $\rho = 0.050$. 


For the laminar state at $a=3.791$, we find $\mathcal{L} = 23.3$
and $\mathcal{C} = 0.79$. The degree centrality $k_v$
follows a distribution with slight bimodality (Fig.~\ref{fig_rho_dist}C). 
The resulting link density approaches its lowest value within
the four considered dynamical regimes ($\rho = 0.040$). 

Finally, for the outer crisis ($a=4$), we obtain $\mathcal{L} = 23.6$ 
and $\mathcal{C} = 0.82$.
The degree centrality $k_v$ displays similar properties as for 
the laminar state, but with higher average values and a resulting
link density of $\rho = 0.046$ (Fig.~\ref{fig_rho_dist}D). 

From the above results, we conclude that complex network measures
applied to a recurrence matrix are indeed sensitive to changes in 
the dynamics. 
The average shortest path length can be considered as an upper bound 
for the phase space distance between two states (in units 
of the threshold value $\varepsilon$). Hence, its average 
value $\mathcal{L}$ can be interpreted as a mean distance, 
which depends on the total diameter and the fragmentation 
of the phase space.
Therefore, $\mathcal{L}$ increases 
with growing phase space of the logistic map (with growing control
parameter $a$). The clustering coefficient $\mathcal{C}$ is able to
detect clustered phase vectors, as they appear in periodic
or laminar dynamics. 
The degree centrality $k_v$ quantifies the phase-space 
density in the direct neighbourhood of a state $v$, while the link density
$\rho$ measures the average phase space density.
Moreover, from the $k_v$ distribution we can infer that the
considered recurrence matrices are not scale-free in the sense of 
the network theory.


Now we calculate $L_{\max}$, $LAM$, $\rho$, $\mathcal{L}$, $\mathcal{C}$, 
and $k_v$ for different values of the control parameter $a$ within the 
range $[3.5,4]$. For each value of $a$, we compute a
time series of length $N=2,000$, and exclude transients
by removing the first $1,000$ values. 

\begin{figure}[hbtp]
\centering \includegraphics[width=.75\textwidth]{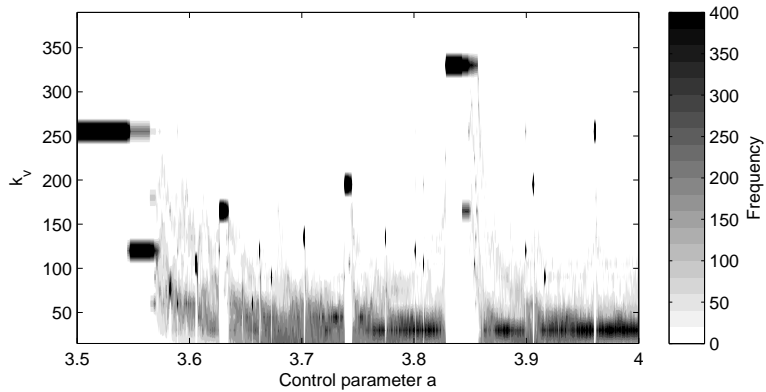}
\caption{Distribution of the degree centrality $k_v$ of the 
logistic map for a range of the control parameter $a$. 
Same parameters as in Figs.~\ref{fig_rho_dist} and 
\ref{fig_measures_a}.}\label{fig_BC_dist_a}
\end{figure}

The RQA measure $L_{\max}$ reveals periodic dynamics by maxima
of its value (Fig.~\ref{fig_measures_a}B). Laminar phases
are clearly detected by $LAM$ (Fig.~\ref{fig_measures_a}C). 
$\rho$ and $\mathcal{C}$ also show maxima
during episodes of periodic dynamics (Figs.~\ref{fig_measures_a}D and F). 
$\rho$ corresponds to the recurrence rate
and confirms previous studies \cite{trulla96}. Its values also depend
on the periodicity during the periodic windows -- the
higher the periodicity, the lower $\rho$. Therefore, period-doublings
cause an abrupt decrease of this measure.
In the periodic regime, neighbours
of a state are equal to the state itself, leading to the largest possible 
clustering coefficient $\mathcal{C} = 1$, and to the shortest possible 
path lengths between
neighbours giving $\mathcal{L} = 1$. However, $\mathcal{L}$
shows a more interesting behaviour.
In our interpretation of a recurrence matrix, $\mathcal{L}$ 
characterises not only the total phase space diameter, but 
also its fragmentation. With respect to the logistic map, 
each time two bands in phase space merge (e.g.~at 
$a=3.5736$ or $a=3.5916$), this does not only lead to an 
increase of the occupied phase space, but also yields a 
merging of formerly disjoint network clusters. As the 
definition of the average path length does not consider 
pairs of points in disconnected clusters, the average 
distance of connected nodes suddenly increases shortly 
before the band merging point as soon as the distance 
between the different bands falls below $\varepsilon$, 
since the clusters then become connected.
This is clearly expressed by jumps in $\mathcal{L}$ (Fig.~\ref{fig_measures_a}E).
The distribution of $k_v$ is discrete in the periodic windows,
which are therefore clearly identifiable (Fig.~\ref{fig_BC_dist_a}). 
Analogous to the link density $\rho$, the location of the maxima of the 
degree distribution in periodic windows is related to the number of
periods, e.g., for period-4 we have $N/4-1=249$, for period-3 
$N/3-1 = 332$ (for a time series length of $N=1,000$). 
The degree distribution $P(k)$ before the band merging point
is broad and reveals higher degrees than after the band merging point, 
which again relates to the connection of the distinct network clusters.
For increased control parameter $a$, $P(k)$ becomes more localised around small degrees, disclosing
the decrease of recurrences due to the increasingly chaotic behaviour
(increasing Lyapunov exponent).

\section{Application to marine dust record}\label{sec:appl}

\begin{figure}[btp]
\centering \includegraphics[width=.85\textwidth]{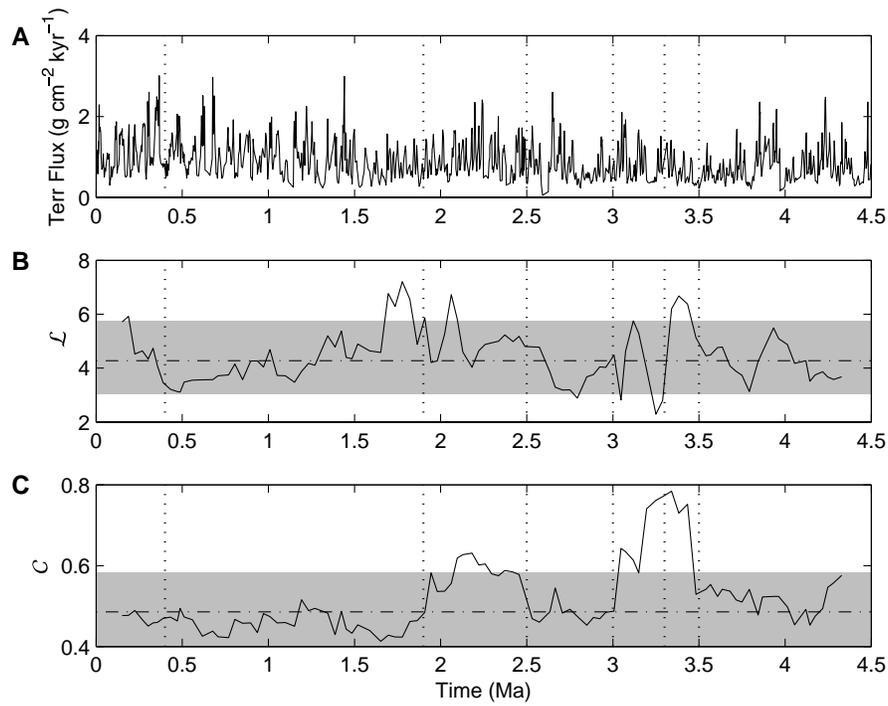}
\caption{(A) Terrigenous dust flux record of ODP site 659, 
and corresponding network measures (B) $\mathcal{L}$ and (C) $\mathcal{C}$.
The dotted lines mark pronounced transitions in the dynamical regime
at 3.5, 3.3, 3, 2.5, 1.9, and 0.4~Ma BP, the dash-dotted line
corresponds to the mean value of the null-model and the
shaded area corresponds to the 90\% confidence bounds.
Same parameters as in Fig.~\ref{fig_dust_RP}, window size 420~ka.
}\label{fig_dust_scalar}
\end{figure}

Long-term variations in aeolian dust deposits are related 
to changes in terrestrial vegetation and are often used as a proxy for
changing climate regimes in the past. For example, marine terrigenous dust records 
can be used to infer epochs of arid continental climate.
In particular, a marine record from the Ocean Drilling Programme
(ODP) derived from a drilling in the Atlantic, ODP site 659, was used to
infer changes in African climate during the last 4.5~Ma 
(Fig.~\ref{fig_dust_scalar}A) \cite{tiedemann1994}. This
time series has a length of $N=1,240$ with an average
sampling time of 4.1~ka. Applying spectral
analysis to these data, it was claimed that the African climate 
has shifted towards arid conditions at 2.8, 1.7 and 1.0~Ma before present (BP) \cite{demenocal1995}. 
These transitions correspond to epochs of different dominant
Milankovich cycles (mid-Pleistocene transition with a ``41~ka world'' between 2.7 and 1.0~Ma BP and a ``100~ka
world'' since about 1.0~Ma BP), the end of the Early Pliocene Warm Period
at about 2.8~Ma BP, and the development of the Walker circulation
around 1.9--1.7~Ma BP \cite{ravelo2004}.
However, a recent thorough investigation of several marine
dust records demonstrated more complex relationships between
vegetational coverage, aeolian transport processes and the
dust flux record \cite{trauth2009}. The analysis revealed
transitions between different regimes of variability, mostly
driven by a variation of the solar irradiation due to different
dominant Milankovich cycles. For example, Trauth et al.~found
an interval of a dominant 100~ka frequency (related to orbital eccentricity)
between 3.2 and 3.0~Ma BP, and of dominant 19--23~ka frequency band (precession)
between 2.3 and 2.0~Ma BP \cite{trauth2009}. The Early Pliocene Warm Period
ended between 3.3 and 2.8~Ma BP with the Pliocene optimum (3.24--3.05~Ma BP)
and the onset of the northern hemisphere glaciation (2.8--2.7~Ma BP)
\cite{mudelsee2005,ravelo2004}, which was intensified during the mid-Pleistocene 
climate shift at 1.0--0.7~Ma BP \cite{stjohn2002}. It has been hypothesised that the latter transition
was connected with a period of strong Walker circulation 
between 1.5--0.5~Ma BP \cite{mcclymont2005}.

We illustrate the capabilities of our recurrence analysis using complex
network measures for the ODP 659 dust flux
record in order to find transitions in the dynamics.
For this purpose, we use a time delay embedding with dimension $m=3$ and delay 
$\tau=2$ (these parameters have been determined by applying the standard procedure
using false nearest neighbours and mutual information \cite{kantz97}). 
The threshold is chosen to preserve a constant recurrence
rate of 5\% (which means that the link density $\rho$ will be 
constant) \cite{marwan2007,schinkel2008}.
In order to study transitions in the dust record, we calculate
the recurrence matrix in moving windows of size 100~time points
(corresponding approximately to 410~ka) and with an overlap of 90\%. 
For the time-scale of the windowed measurements, we use the mid-point of the window.
Note that the time-scale is not equidistant (equidistant time-scale is
not necessary for our network approach). On average, the sampling
time is 4.1~ka with a standard deviation of $\sigma=2.7$~ka. Compared
to the long (geological) scale this deviation is still rather small.
However, the application of linear methods often requires equidistant
time-scales.

\begin{figure}[tbp]
\centering \includegraphics[width=.7\textwidth]{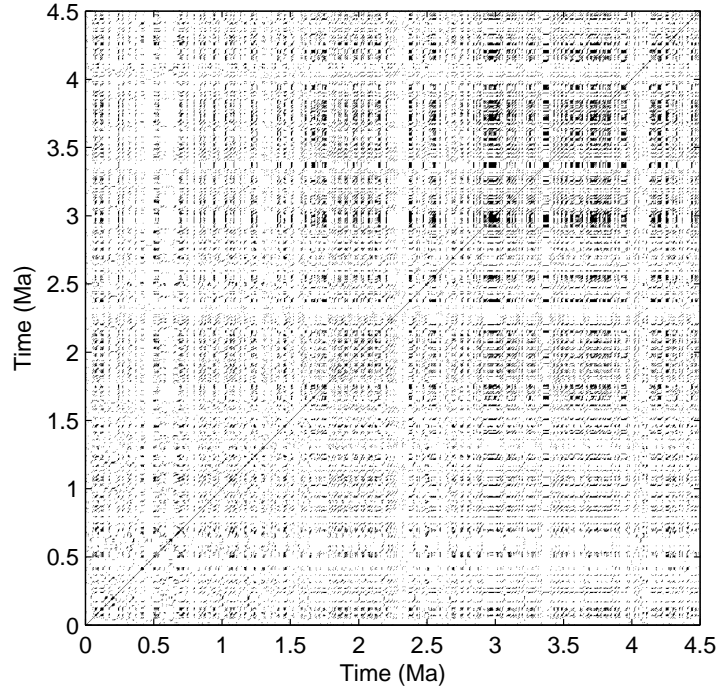}
\caption{Recurrence plot of the terrigenous dust flux record of ODP site 659.
Parameters are $m=3$, $\tau=2$, $\varepsilon$ is chosen such that 
$\rho = 0.05$, phase space distances are measured using maximum norm.
}\label{fig_dust_RP}
\end{figure}

We will apply a simple statistical test in order to test the
null-hypothesis to see whether the network characteristics
at a certain time differs from the general network characteristics.
In order to create an appropriate null-model, we use the following
approach. In contrast to the RQA measures, where the time-ordering is
important requiring a more advanced approach for a statistical
test \cite{schinkel2009a}, for the network measures we can
simply randomise the time series: we randomly draw 100 values
(corresponding to the window size of 100 points) from the time series
and then calculate the RP and the network measures from this sample.
By repeating this 10,000 times we get a test distribution for the
measures $\mathcal{L}$ and $\mathcal{C}$ and estimate its 0.05 and
0.95 quantiles that may be interpreted as the 90\% confidence bounds.

The RP of the dust data depicts a rather homogeneous recurrence
structure, interrupted only by rather small bands of sparse
recurrence point density (Fig.~\ref{fig_dust_RP}). Such sparse areas 
mark epochs of more frequently occurring extreme or rare events recorded by the
marine dust data series. On the small-scale we find mostly very short 
diagonal lines, expressing the high variability and fast change of 
the states (with respect to the geological time-scale). Between 4.0 and 3.0~Ma
and around 2.0~Ma~BP, longer diagonal lines appear. Moreover, between 4.5
and 3.0~Ma~BP, we find an increased number of vertical/ horizontal
lines, indicating different dynamics than at other times.

\begin{figure}[hbtp]
\centering \includegraphics[width=.75\textwidth]{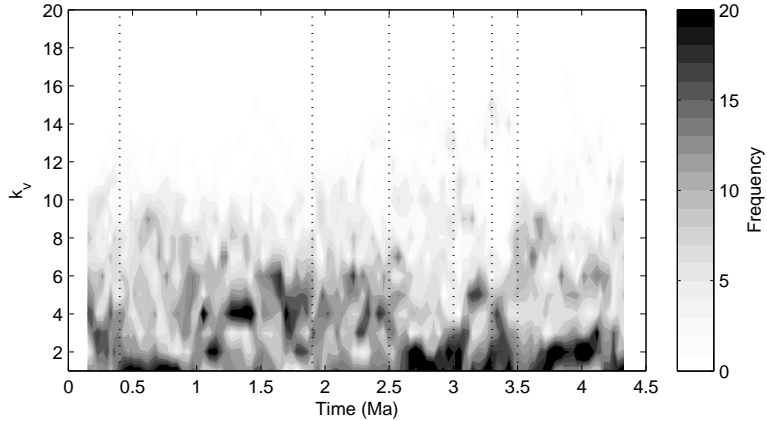}
\caption{Degree centrality $k_v$ of terrigenous dust flux record of ODP 
site 659. Same parameters as in Figs.~\ref{fig_dust_RP} and
\ref{fig_dust_scalar}. 
}\label{fig_dust_distr}
\end{figure}

The global network measures $\mathcal{L}$
and $\mathcal{C}$ also depict a distinct
variability (Fig.~\ref{fig_dust_scalar}B and C). 
$\mathcal{L}$ reveals epochs
of significantly higher values between 3.5 and 3.3, $\sim 2.1$, 
1.9--1.8, and after 0.4~Ma BP. Around 3.3, 2.0 and 1.9~Ma~BP the RP
exhibits sudden drops of $\mathcal{L}$ within a period of,
in general, higher values. $\mathcal{C}$ discloses
epochs of increased values between 3.5 and 3.0~Ma as well as between
2.5 and 2.0~Ma~BP. Between 4.5 and 3.5~Ma, 3.0 and 2.5~Ma, and 
1.0 and 0.4~Ma~BP, 
the degree centrality possesses mostly small
values, whereas between 2.5 and 1.0~Ma and after 0.4~Ma~BP,
it has larger values (Fig.~\ref{fig_dust_distr}). 



With respect to the previously known results, we 
conclude that $\mathcal{C}$ identifies the epochs 
of more dominant Milankovich cycles (between 3.2 and 3.0~Ma 
and 2.3 and 2.0~Ma BP). 
$k_v$ is increased in these periods, but also
exhibits increased values for the period between 2.5 
and 1.0~Ma BP. Note that the 3.5-3.0~Ma BP period is 
related to the intensification of the Northern hemisphere 
glaciation \cite{berger1994}.
In contrast,  $\mathcal{L}$ reveals 
transitions in climate dynamics on a different time-scale. 
Maxima of this measure tend to appear at the onset of 
changes in $\mathcal{C}$. Whereas  
$\mathcal{C}$ reveals the changed dynamics, 
$\mathcal{L}$ is sensitive to the transition 
periods, which is consistent with our results near the 
band-merging points of the logistic map. 
The increase of $\mathcal{L}$ at $\sim 3.4$, $\sim 3.1$, 
1.9--1.8, and 0.4~Ma BP may also be related to the 
intensification of glaciation. However, the detected 
transitions are associated with different and more subtle 
dynamical changes, and not simply just an 
intensification of a certain Milankovich cycle.

\section{Conclusions}

We have linked the recurrence matrix with the adjacency matrix 
of a complex network, and have proposed the direct application of 
the corresponding network measures to the recurrence matrix.
We have discussed the link density, degree centrality,
average path length and clustering coefficient 
in some detail. In particular, the latter two complex network
measures have no direct counterpart in recurrence
quantification analysis and give additional insights
into the recurrence structure of dynamical systems. In a
further study, we have outlined the link between the complex network measures
and the properties of the phase space trajectory of dynamical systems \cite{donner2009netw}.

By applying our novel approach to the logistic map, we have illustrated
the ability of the proposed measures to distinguish between the
different dynamical regimes and to detect the corresponding transitions.
Moreover, we have used our approach to investigate
a marine climate proxy record representing the climate variability
over Africa during the last 4.5~Ma. The different measures
highlighted various transitions in the recurrence structure
and, hence, in the dynamics of the studied climate system.
By applying the recurrence
approach and complex network measures, we were able to identify
more subtle transitions than those that were previously reported 
from using linear approaches, like 
power spectral analysis \cite{demenocal1995}, linear trend
detection \cite{mudelsee2005}, Mann-Whitney or 
Ansari-Bradley tests \cite{trauth2009}.
In addition, our proposed approach to detect transitions on the
basis of time series does not require equidistant
time-scales, as would be necessary for most other
known techniques. From the network point of view, the 
recurrence plot approach can deliver a potential
measure of information exchange in time series of
complex systems \cite{west2008,donner2009netw}.

In the future, recurrence plots and their complex network interpretation 
will allow for further fruitful and natural transfer of 
ideas and techniques from complex network theory to time 
series analysis (and vice versa).

\section{Acknowledgements}
This work was partly supported by the German Research Foundation 
(DFG) project He 2789/8-2, SFB 555 project C1, and the Japanese Ministry for 
Science and Education.

\bibliographystyle{elsarticle-num-names}
\bibliography{mybibs,rp,cn,geo}

\end{document}